\documentclass[preprint,pre,eqsecnum]{revtex4-1}
\usepackage[utf8]{inputenc}
\usepackage{amsmath}
\usepackage{amssymb}
\usepackage{lipsum}
\usepackage{bm}
\usepackage{graphicx}
\usepackage[dvipdf]{epsfig}
\usepackage{float}
\usepackage{wrapfig}
\usepackage{mathtools}
\usepackage{hyperref}
\usepackage[dvipsnames]{xcolor}

\usepackage{ulem}  %\sout{stuff to eliminate}
\usepackage{cancel} %\cancel{stuff to eliminate}

\begin{document}
\title{The propagation of infection fronts in spatially distributed 
compartment models in epidemiology}

\author{Joseph Rudnick$^{1}$, David Jasnow$^{2}$, and Jorge 
Vi\~nals$^{3}$}
\affiliation{$^{1}$ Department of Physics and Astronomy, University of
California Los Angeles, CA 90095, $^{2}$ Department of Physics and Astronomy,
University of Pittsburgh, Pittsburgh, PA 15260, $^{(3)}$ School of Physics and
Astronomy, and Institute for Health Informatics, University of Minnesota,
Minneapolis, MN 55455, USA}

\begin{abstract}
Spatio-temporal extensions of familiar compartment models for disease transmission incorporating diffusive behavior, or interactions between individuals at separate locations, are explored. The models considered have the character of reaction-diffusion systems, which allow familiar techniques to be applied. The focus is largely on the appearance of soliton-like moving fronts that spread infection to previously uninfected regions.  
Near threshold dynamical critical behavior and a degree of universality are revealed. Extending two of the models to include a simple nonlinearity in the strength of the binary interaction between a susceptible individual and an infected one, we find the possibility of static coexistence between spatial regions with different levels of infection and an analogy with first-order transitions in thermodynamics.

\end{abstract}

\date{\today}

\maketitle

\section{Introduction}

The so called compartmental models in epidemiology follow the foundational research of Kermack and McKendrick \cite{re:kermack27}, who formulated kinetic equations to describe the temporal evolution of an epidemic in terms of various subpopulations and their interactions. For example, subpopulations can refer to individuals susceptible to a disease, those exposed to it, infected, or recovered \cite{re:anderson79,re:murray01,re:li18}. Initially space-independent, the solutions of these models provided key insights into epidemic progression mechanisms, and into the existence of steady states that were either disease free, or associated with persistent infection. The models also focused on the characteristic rates of the various stages of disease progression, and led to predictions of the characteristic time scales and infection levels that followed. The emergence of SARS-CoV-2 has spurred new research into compartment models of disease spread, specially given the wealth of fine grained, time and space dependent, empirical data. Generalizations of compartment models in this case include, for example, the consideration of fine grained compartments \cite{re:parolini21,re:yan21}, a cell based network that also incorporates subpopulation fluxes to mimic population mobility \cite{re:letreut21}, or the addition of delays relating the newly infected and the newly recovered \cite{re:huang21}

More generally, a key development in the study of population dynamics is the work of Skellam \cite{re:skellam51}, which was inspired by earlier work of Fisher \cite{re:fisher37}  on evolutionary genetics, and who introduced diffusive processes (Brownian motion) to model species dispersal in Europe. The concept of diffusive dispersal has been widely applied in Biology \cite{re:belgacem01,re:fife13}, including in the study of disease spreading and a theory of epidemics \cite{re:hoppensteadt12,re:postnikov07,re:abdelheq19}. Beginning with the work of Noble \cite{re:noble74}, attention was paid to the details of the geographic spread of epidemics by allowing and explicitly modeling the spatio-temporal evolution of the subpopulations involved. Therefore spatially dependent generalizations of compartment models play an important role in describing the spread of disease as they intend to capture the effect of mobile disease carriers across geographical regions. Communication occurs through migration \cite{re:noble74}, or through contacts between infected and susceptible individuals, and therefore predominantly between adjacent geographical areas \cite{re:naether08}. This research highlighted the possible existence of traveling fronts separating regions with high prevalence of the disease from regions with healthier populations. Results were obtained on the conditions for fronts to develop and propagate, as well as their speed. Noteworthy is the analysis of K\"all\'en et al. \cite{Kallen} of a model of the spread of rabies in Europe. The model studied is a variant of the so called SI model (definitions of the various models are given below), and it was shown to lead to a propagating front of infection, with a velocity which is substantially equivalent to the results which we present below. We mention here the general limitation that the spatial coupling or correlation among compartments in these models is rather simple, as they are generally restricted to a small spatial neighborhood of any given segment of the population, and the dynamics follows rather idealized assumptions (e.g., simple diffusion). 

More recently, compartment models have been extended to include complex networks that incorporate a large number of distinct nodes (generalized compartments). The nodes form an interconnected network with some specified topology, and are allowed to interact in more general ways than simple geographical proximity \cite{re:rock14,re:verdasca05,re:eksin21}. Recent examples of application of the methodology include the study of waves in the spread of influenza \cite{re:viboud06}, an empirical determination of a network of social contacts, and its consequences on disease incidence \cite{re:mossong08},  or the spread of SARS-CoV-2 associated with a number of connectedness and mobility indices \cite{re:holtz20,re:paltiel20,re:letreut21}. A modified SIR(S) model has been introduced to study the effect of changing climate on basic infectious disease reproduction number, and modeled heterogeneity by introducing a set of coupled compartments into the model \cite{re:saad20}. Many current SARS-CoV-2 studies involve contact networks with a very large number of nodes, on the order of one million (the population of a typical metropolitan area) \cite{re:nande21}. The role of explicit delay terms in the differential equations defining a compartment model has also been considered to describe SARS-CoV-2 outbreaks in Illinois in Ref. \cite{re:wong20}. These methods are considered predictive in that model parameters are estimated by statistical fits to past empirical data, and then models can be propagated into the future in order to predict the evolution of an epidemic. Finally, we mention results concerning front propagation in more general models that consider nonlocal interaction kernels and diffusivity in SEIR compartment models. The conditions for the existence of traveling solutions in the model have been established \cite{re:wu18}, and later extended to include the effects of temporal delays in the interaction kernels \cite{re:wu21}.

The models that we analyze below are written in terms of coarse grained variables describing the number of individuals belonging to a certain subpopulation. For example, and following standard convention, we denote by $S(x,t)$ the number density of susceptible individuals in an element of volume around $x$ at time $t$, $E(x,t)$ is the number of individuals who have been exposed to the infectious disease, leading to a number, $I(x,t)$, of infected and $R(x,t)$ recovered individuals. The various models are described in their corresponding section, but they all involve rate functions describing transitions among the subpopulations, and either spatial diffusion of individuals belonging to the various subpopulations, or space-dependent disease transmission among them depending on the model. As a consequence of their definition, all the models studied here belong to the class of reaction-diffusion equations, and we therefore use the tools of bifurcation theory and nonlinear dynamics to analyze solutions of interest. We restrict our analysis to the simpler case of one dimensional spatial variations, although the models are readily generalized to higher dimensions. We find in all cases that the evolution of some of the dynamical variables satisfies the Fisher equation (Appendix \ref{sec:fkpp}) \cite{re:fisher37,re:kolmogorov37} \emph{near onset} of infection propagation. This connection allows us to find analytic expressions for the shape of the propagating front and its velocity in some cases, whereas in others we proceed numerically to confirm the existence of fronts and obtain their velocity.

In Section \ref{sec:noble} we begin by considering a simple SI model that includes only two populations, susceptible and infected, as first studied by Noble \cite{re:noble74}. This is a model that addresses the evolution of a disease that spreads through a population but that, at asymptotically long times, there are no infected individuals remaining. We show in this case that a wave of infection exists originating from an initial condition in which a few individuals are assumed to be infected in some small region. If the infection rate is larger than the recovery rate, a wave of infection sweeps across the entire system. The wave envelope has constant shape as it propagates as a front of constant velocity, which we compute numerically. The same method of analysis is used next to describe the SIR model with vitality, with similar conclusions when the rate of infection is larger that the rate of recovery plus the rate of population renewal. The same results obtain for the more complex SEIR model, which we show reduces, near threshold,  to the Fisher-Kolmogorov model for each of the subpopulations. We also look at a model which ignores spread through diffusion of individual populations but takes into account non-locality of the infection process. Finally we examine extended SIR and SEIR models in which the infectivity rate coefficient is assumed to depend on the number of infected individuals. The added nonlinearity allows bistability between two fixed points, one featuring persistent infection, the other disease free. In this case we find a unique front solution separating both fixed points, and compute the front velocity for a range of parameter models.

\section{Noble's SI model} \label{sec:noble}

In a seminal paper \cite{re:noble74}, J. V. Noble investigated a simple epidemiological model for the propagation of an infection, which spreads by means of a moving front. The model divides individuals into two populations, consisting of those who are disease-free and susceptible to the infection ($S$) and those who have been infected ($I$). The ultimate fate of the members of the $I$ population is to be removed from the system, either through death or by virtue of acquired immunity.     This model is appropriate to an epidemic such as the bubonic plague in the pre-antibiotic era which rapidly  sweeps through a population, causing most of the deaths that occur in the time-frame of its duration. The one-dimensional equations for this system are {taken to be}
\begin{eqnarray}
\frac{\partial S(x,t)}{\partial t} & = & D \frac{\partial^2 S(x,t)}{\partial x^2} - \beta S(x,t) I(x,t) \label{eq:outline1} \\
\frac{\partial I(x,t)}{\partial t} & = &  D \frac{\partial^2 I(x,t)}{\partial x^2} + \beta S(x,t) I(x,t) - \mu I(x,t) \label{eq:outline2}
\end{eqnarray}
The population of susceptible individuals, $S(x,t)$, changes through diffusive motion, quantified by the coefficient $D$, and attrition due to infection, the rate being determined by the extent of contact between the two populations and an infectivity coefficient, $\beta$. The infected population, $I(x,t)$ changes as a result of the process of infection, diffusion and the rate at which individuals are removed through death or the acquisition of immunity, governed by the coefficient $\mu$.  

We start with the case of a uniform (spatially mixed) set of populations, in which the model can be analytically solved with the use of quadratures, and which guides the exploration of the transition between the conditions under which the infection dies out locally and those  in which it spreads and develops into an epidemic.

\subsection{The equations assuming uniform populations} \label{sec:uniform}

In this case the spatial dependence in Eqs.  (\ref{eq:outline1}) and (\ref{eq:outline2}) can be ignored.  They then reduce to
\begin{eqnarray}
\frac{dS(t)}{dt} & = & - \beta S(t) I(t) \label{eq:outline3} \\
\frac{dI(t)}{dt} & = & \beta S(t) I(t) - \mu I(t) \label{eq:outline4}
\end{eqnarray}
Dividing (\ref{eq:outline4}) by (\ref{eq:outline3}), we end up with the single differential equation
\begin{equation}
\frac{dI}{dS} = -1+ \frac{\mu}{\beta S} \label{eq:outline5}
\end{equation}
Integrating this equation up, we find
\begin{equation}
I=I_0 -(S-S_0) + \frac{\mu}{\beta} \ln(S/S_0) \label{eq:outline6}
\end{equation}
where $I_0$ and $S_0$ are the initial values of the infected and susceptible populations. For further discussion of the analysis of this model see \cite{Brauer}.

\begin{figure}[htbp]
\begin{center}
\includegraphics[width=5in]{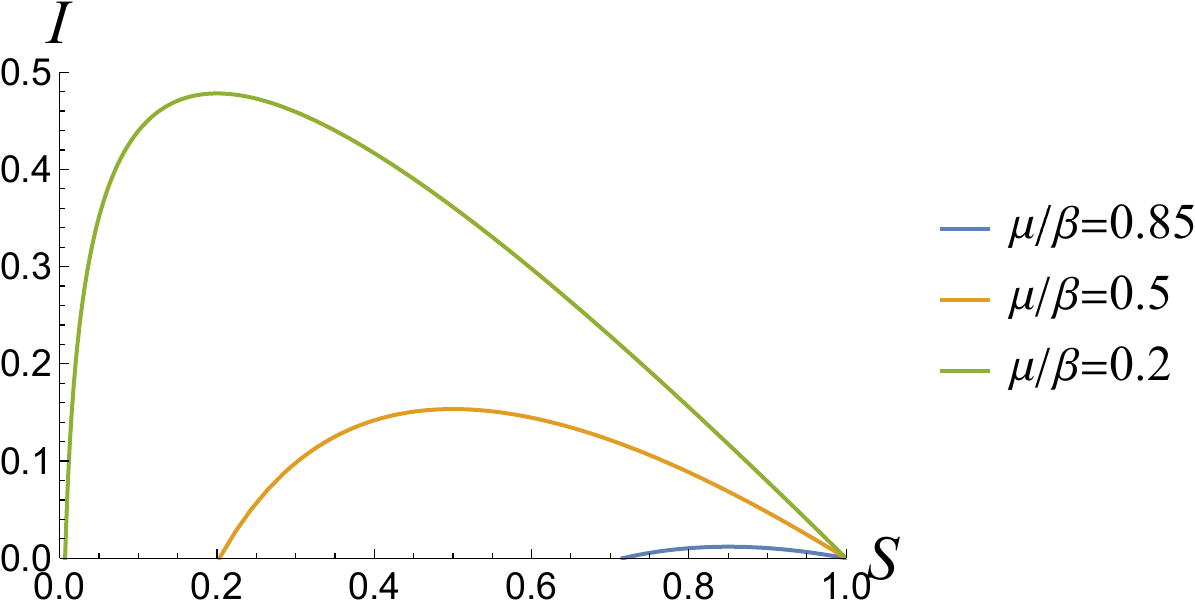}
\caption{A plot of $I(t)$ versus $S(t)$ for various values of $\mu/\beta$. The initial value of $S$ is 1 and the initial value of $I$ is 0 The time flow for all plots is to the left, in the direction of decreasing $S(t)$. The plots terminate when $I(t)=0$, at which point there is no change in either $S(t)$ or $I(t)$. The points at which a curve intersects the horizontal axis are both  steady states. The state at the right is unstable and the state at the left is stable. }
\label{fig:SandI}
\end{center}
\end{figure}

Note the properties of the equations above. First, given (\ref{eq:outline3}), $S(t)$ is monotonically decreasing. In light of this fact, the change with $t$ of $I(t)$ for small values of $t$ is controlled by the sign of the right hand side of (\ref{eq:outline5}); if it is negative $I(t)$ increases with $t$, and if it is positive $I(t)$ decreases with $t$.  Given the form of (\ref{eq:outline5}) and the monotonic behavior of $S(t)$ we know that eventually $I(t)$ will decrease with time.  Whether it ever increases is determined by the values of $\beta$, $\mu$ and $S_0$. When $\beta > S_0 \mu$, $I(t)$ decreases monotonically with $t$, and when $\beta < S_0 \mu$, there is an initial increase with time before the inevitable decline of $I(t)$. As background for the analysis to follow, we set the initial conditions to $I_0=0$ and $S_0=1$. This corresponds to an uninfected system. However, we will use it as the limit of a system with an extremely small infected population. The parametric solution (\ref{eq:outline6}) then becomes
\begin{equation}
I= 1-S+ \frac{\mu}{\beta} \ln(S) \label{eq:parametric1}
\end{equation}
At $S=1$, the slope of the curve---see Fig. \ref{fig:SandI}---is
\begin{equation}
\left.\frac{dI}{dS} \right|_{S=1} = \frac{\mu}{\beta} -1 \label{eq:parametric2}
\end{equation}
If $\beta>\mu$, the $I$ vs. $S$ curve has a negative slope.  Given the time dependence of $S(t)$ this means $I(t)$ increases from its initial value, {(near)} zero, with time. If $\beta < \mu$ there is no such growth, and the infection never spreads.

\begin{figure}[htbp]
\begin{center}
\includegraphics[width=4in]{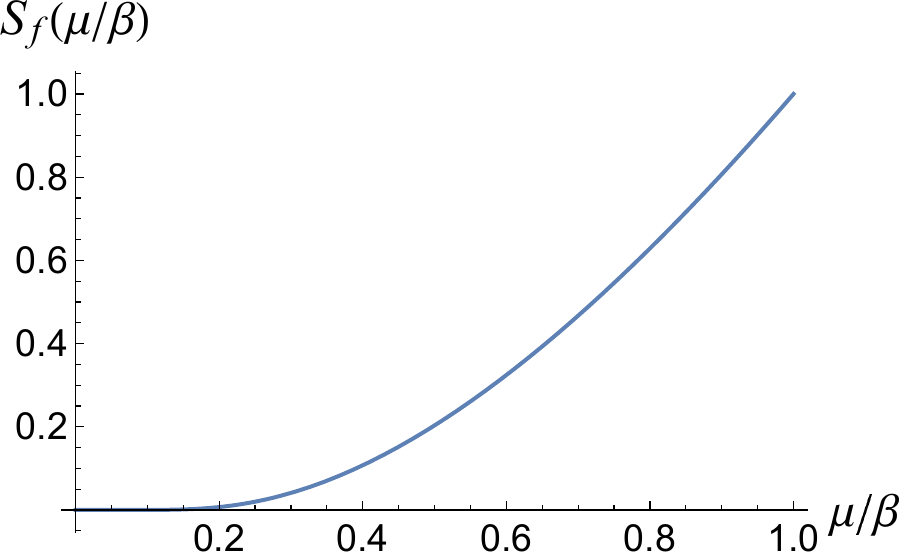}
\caption{Plot of $S_f(\mu/\beta)$.}
\label{fig:Sfinal}
\end{center}
\end{figure}

As for the late-time behavior of $S(t)$ and $I(t)$, the last term on the right hand side of (\ref{eq:outline6}) guarantees that there will be a value of $t$ for which $I(t)$ passes through zero. At that time, the absence of an infected population causes the time derivative of $S(t)$ to vanish, while Eq. (\ref{eq:outline5}) guarantees that  $I(t)$ is also stationary, A dynamically stable steady state has thus been reached. The value of $S$ at this point is determined by the ratio  $\mu/\beta$. There is an explicit formula for this value, obtained by solving (\ref{eq:parametric1}) with the left hand side set equal to 0. {One finds} for that value
\begin{equation}
S_f(\mu/\beta)= -\frac{\mu }{\beta} W\left(-\frac{\beta }{\mu} e^{-\frac{\beta }{\mu }}\right)
\label{eq:Lambert}
\end{equation}
where $W(x)$ is the Lambert, or product log, function.\cite{Lambert} Figure \ref{fig:Sfinal} shows how the final value of $S$ depends on the ratio $\mu/\beta$. 

Finally, the full solutions of Eqs.(\ref{eq:outline1}) and (\ref{eq:outline2}) can be numerically completed by quadratures. Substituting the right hand side of (\ref{eq:parametric1}) for $I$ in (\ref{eq:outline3}), an implicit solution for $S(t)$ is obtained by straightforward integration. Then, inserting that solution into the right hand side of (\ref{eq:outline4}) we recover $I(t)$ by numerical integration.

\subsection{The spread of the infection as a propagating front:  determination of the front's velocity} \label{sec:velocity1}

\begin{figure}[htbp]
\begin{center}
\includegraphics[width=5in]{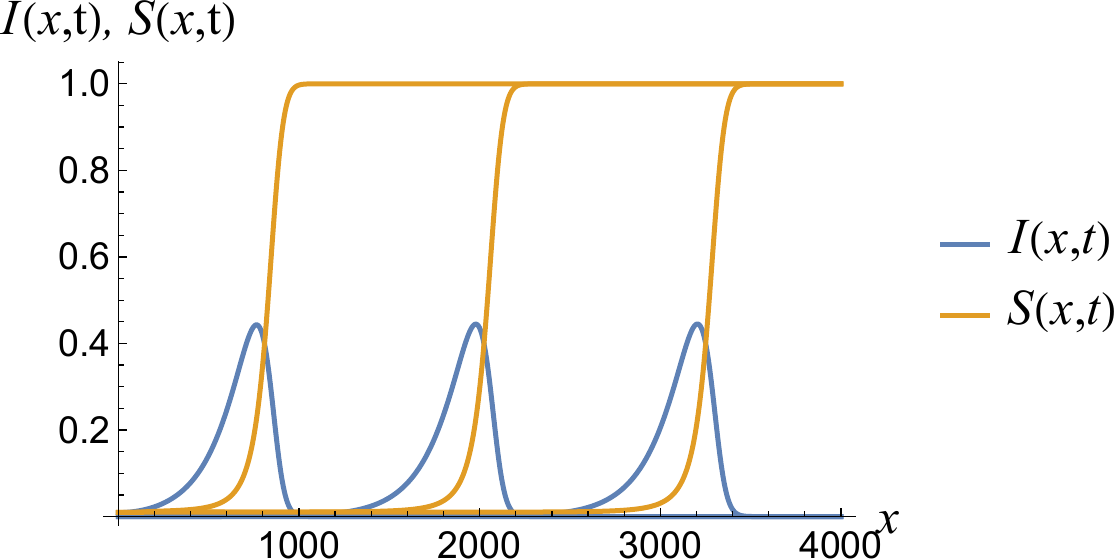}
\caption{The solutions of the Noble equations Eqs. (\ref{eq:outline1}) and (\ref{eq:outline2}) for  $I(x,t)$ and $S(x,t)$ plotted for three equally spaced times. The coefficients in those equations are $D=300$, $\beta=1$ and $\mu=0.2$. The fronts propagate from the left to the right. See also corresponding plots in Noble \cite{re:noble74}. As the advancing front sweeps through the system the initial state of $S=1, I=0$  transforms to a final steady state in which $I=0$ and $S$ is reduced to a smaller, but non-zero, value.}
\label{fig:fronts1}
\end{center}
\end{figure}

According to  numerical evidence, after transients have died off a spreading infection propagates as a front which retains its shape as it translates; i.e. it is soliton-like. Figure \ref{fig:fronts1} displays the results of the numerical solution of Eqs. (\ref{eq:outline1}) and (\ref{eq:outline2}). 

\begin{figure}[htbp]
\begin{center}
\includegraphics[width=4in]{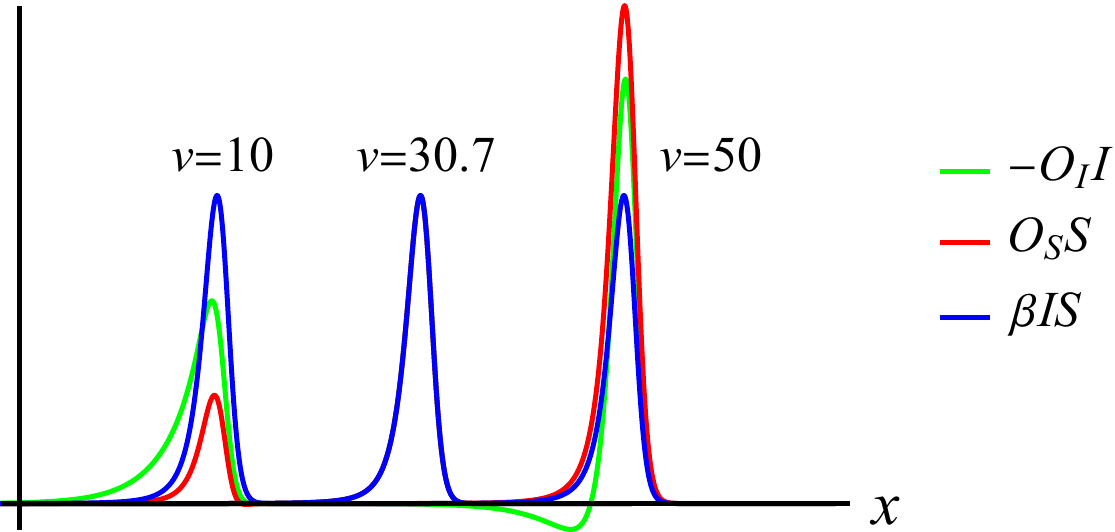}
\caption{The three functions in (\ref{eq:outline24}), plotted for three assumed front velocities, $v$. The three sets of curves are offset along the $x$ axis for clarity of presentation. For only one of the three velocities---the actual propagation velocity of the front---do the three functions coincide. The coefficients are the same as in Fig. \ref{fig:fronts1}.}
\label{fig:velocity}
\end{center}
\end{figure}

If the propagation of the fronts is indeed soliton-like, then we can write 
\begin{eqnarray}
S(x,t) &=& S_s(x-vt)  \label{eq:outline20} \\
I(x,t) & = & I_s(x-vt) \label{eq:outline21}
\end{eqnarray}
where the subscript $s$ indicates soliton-like structure. If we define
\begin{eqnarray}
O_S& = & v \frac{d}{dx} + D \frac{d^2}{dx^2}  \label{eq:outline22} \\
O_I & = & v \frac{d}{dx} + D \frac{d^2}{dx^2} + \mu \label{eq:outline23}
\end{eqnarray}
the equations for $S_s$ and  and $I_s$ that follow from (\ref{eq:outline1}) and (\ref{eq:outline2}) can be reduced to 
\begin{equation}
O_S  S_s(x)= -O_I I_s(x) = \beta S_s(x) I_s(x) \label{eq:outline24} 
\end{equation}
This allows us to utilize the numerical solution of equation set (\ref{eq:outline1}) and (\ref{eq:outline2}) to extract the value of the velocity, $v$ of propagation of the contagion. An example of the determination of the {propagation} velocity  of the fronts is shown in  Fig. \ref{fig:velocity}. The three functions coincide for only one of the three  assumed velocities ($v=30.7$).

 \subsection{The transition between a dissipating and a propagating infection}
 As in the case of the uniform system considered in Section \ref{sec:uniform}, the evolution of the system described by the spatiotemporal equations (\ref{eq:outline1}) and (\ref{eq:outline2}) changes dramatically as the ratio $\beta/\mu$ passes through its threshold value. When $\beta/\mu $ is larger than the value of the threshold, the infection propagates through the system as a soliton-like front as noted above, while if the ratio is less than the threshold the infection remains localized and eventually dies off. This fundamental change in the nature of the contagion is a dynamical phase transition, and like many phase transitions, it displays distinctive properties, including diverging space and time scales governed by critical exponents.

\subsection{The uniform population just above threshold}

Referring back to Section \ref{sec:uniform}, particularly Eq. (\ref{eq:outline6}), we can expand about the threshold for increasing infected population. We set
$I_0=0{+}$, $S_0=1$, $S=1-\Delta S$, with $\Delta S \ll 1$ and $\mu/\beta = 1-\epsilon$ with positive $\epsilon \ll 1$ {measuring the deviation from threshold}. (Note that $I_0$, a positive infinitesimal, and $S_0$ will refer to a time in the distant past.) In this case, expanding to second order in $\Delta S$ we find
\begin{equation}
I= \epsilon \Delta S - \frac{\Delta S^2}{2} \label{eq:outline7}
\end{equation} 
We now turn to (\ref{eq:outline3}),  the equation for the time dependence of $S(t)$. Using the expansion above, we find
\begin{equation}
\frac{d \Delta S}{dt} = \beta (\epsilon \Delta S - \Delta S^2/2) \label{eq:outline8}
\end{equation}
which is readily integrated up. We find
\begin{equation}
\Delta S(t) = \frac{2 \epsilon e^{ \beta \epsilon t}}{1+e^{ \beta \epsilon t}} \label{eq:outline9}
\end{equation}
If we plug this form into Eq. (\ref{eq:outline7}), we find 
\begin{equation}
I(t) = \frac{2 \epsilon^2 e^{ \beta \epsilon t}}{(1+e^{ \beta \epsilon t})^2} \label{eq:outline10}
\end{equation}
The net change in the susceptible population as a result of the propagating infection vanishes as {$\epsilon \rightarrow 0$} 
while the maximum value of the infected population over the course of the infection scales as {$\epsilon^2$}. 
Furthermore, the length of time over which the changes arising from the infection take place diverges as {$\epsilon^{-1}$}.

\subsection{A representation of the SI system immediately above threshold}  \label{sec:SIthreshold}
We now know three things about the solution to the SI equations in the {\textit{uniform system}} near onset of propagating fronts. First, the difference between $S(t)$ and its uninfected value is of order $\epsilon$.  Second, we know that the magnitude of the infected population goes as $\epsilon^2$. Third, we know that the time dependence slows down as $\epsilon \rightarrow 0$, the time scale going as $1/ \epsilon$. 

We now set the initial susceptible population infinitesimally smaller than one. We then set the infectivity coefficient $\beta$ equal to one and the death rate $\mu$ equal to $1- \epsilon$. Then, we re-parameterize the populations as follows
\begin{eqnarray}
S(x,t) &=& 1- \epsilon + \epsilon \sigma(\epsilon t, \sqrt{\epsilon} x) \label{eq:outline11} \\
I(x,t)  & = & \epsilon^2 \iota(\epsilon t, \sqrt{\epsilon} x) \label{eq:outline12}
\end{eqnarray}
The particular form of the presumed $x$-dependence is motivated by the desire to assure that the diffusion 
 terms behave in the same way as the time derivative contributions as $\epsilon \rightarrow 0$. 
Substituting into Eqs. (\ref{eq:outline1}) and (\ref{eq:outline2}) and extracting the leading order terms---which means we can neglect the difference between $S(x,t)$ and 1 on the right hand side of 
{the scaled} version of Eq. (\ref{eq:outline1})---we end up with the following two equations describing the behavior of the two functions $\sigma$ and $\iota$
\begin{eqnarray}
\frac{\partial \sigma( \tau, \chi)}{\partial \tau} & = & D \frac{\partial^2 \sigma( \tau, \chi)}{\partial \chi^2} - \iota(\tau, \chi) \label{eq:outline13} \\ 
\frac{\partial \iota( \tau, \chi)}{ \partial \tau} & = & D \frac{\partial^2 \iota( \tau, \chi)}{\partial \chi^2} + \iota(\tau, \chi) \sigma(\tau, \chi) \label{eq:outline14}
\end{eqnarray}
Here,
\begin{eqnarray}
\tau & = & \epsilon t \label{eq:outline15} \\
\chi & = & \sqrt{\epsilon} x  \label{eq:outline16}
\end{eqnarray}

The equation set (\ref{eq:outline13}) and (\ref{eq:outline14}) can be readily solved numerically. 
\begin{figure}[htbp]
\begin{center}
\includegraphics[width=4in]{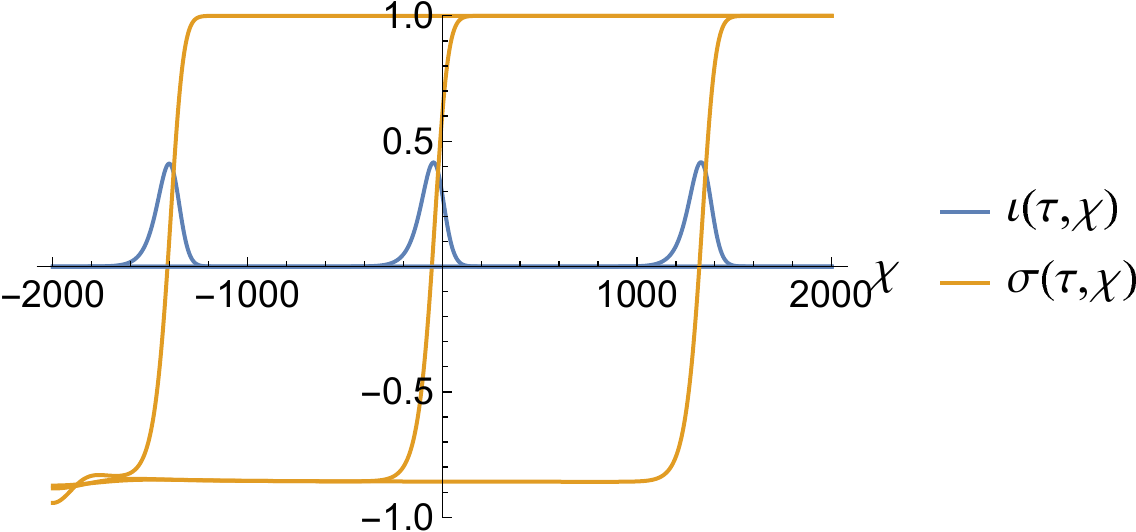}
\caption{The solutions to Eqs. (\ref{eq:outline13}) and (\ref{eq:outline14}) for three equally-spaced values of the rescaled time $\tau$. The diffusion constant $D$ has been set equal to 300.}
\label{fig:populations}
\end{center}
\end{figure}
Figure \ref{fig:populations} shows the result of implementing those equations for the case $D=300$. This is clear evidence for soliton-like behavior. The behavior of the curves for $\sigma(\tau, \chi)$ at the far left reflects the relaxation of vestiges of initial conditions to the stable steady state at very large values of $\tau$.  We can check for soliton-like behavior by assuming that $\sigma(\tau, \chi)$ has the form $\sigma_s(\chi-v^{\prime} \tau)$, and similarly for $\iota$. , 
i.e.,
\begin{eqnarray}
\sigma( \chi, \tau) & = & \sigma_s( \chi- v^{\prime} \tau) \label{eq:scalingsigma} \\
\iota( \chi, \tau) & = & \iota_s( \chi- v^{\prime} \tau) \label{eq:scalingtau}
\end{eqnarray}

Then, (\ref{eq:outline13}) and (\ref{eq:outline14}) take the form
\begin{eqnarray} 
v^{\prime} \sigma_s^{\prime}(\chi-v^{\prime}\tau) +D \sigma_s^{\prime \prime}(\chi-v^{\prime} \tau)&=& \iota_s(\chi-v^{\prime}\tau) \label{eq:outline17} \\
v^{\prime} \iota_s^{\prime} (\chi-v^{\prime} \tau) + D \iota_s^{\prime \prime}(\chi-v^{\prime} \tau) & = & - \iota_s(\chi-v^{\prime} \tau) \sigma_s( \chi - v^{\prime}\tau) \label{eq:outline18}
\end{eqnarray}
Both equations will be satisfied with the correct choice of the {selected} velocity $v^{\prime}$. The velocity can be determined inserting numerical solutions of  (\ref{eq:outline13}) and (\ref{eq:outline14}) for a fixed value of $t$ into the right and left hand sides of (\ref{eq:outline17}) and (\ref{eq:outline18}) and then determining the value of $v^{\prime}$ for which those equations are satisfied. Numerics show that the proper choice for the velocity, $v^{\prime}$, is 
\begin{equation}
v^{\prime} \simeq 2\sqrt{D} \label{eq:outline19}
\end{equation}
Given the relationships (\ref{eq:outline15}) and (\ref{eq:outline16}) this corresponds to a velocity in the original coordinates scaling as
\begin{eqnarray}
v & = & \epsilon^{1/2} v^{\prime} \nonumber \\
& \simeq & 2 \sqrt{\epsilon D} \label{eq:noblev}
\end{eqnarray}

.

\section{SIR model with vitality} \label{sec:SIR}

 It is possible to amend the Noble SI equation set (\ref{eq:outline1}), (\ref{eq:outline2}) by adding an equation for the 
 dynamics 
 of the population of ``removed'' individuals, $R(x,t)$. 
 This new equation is
\begin{equation}
\frac{\partial R(x,t)}{\partial t} = D \frac{\partial^2R(x,t)}{dx^2} + \mu I(x,t) \label{eq:Nsir}
\end{equation}
The resulting equation set is an SIR model for the evolution of a contagion.  No new behavior results from the expansion of the model to incorporate the population of individuals who have contracted the infection and either died or survived with protection against reinfection. However, there is an interesting variant of the model consisting of Eqs.  (\ref{eq:outline1}), (\ref{eq:outline2}) and (\ref{eq:Nsir}). This is the SIR model \textit{with vitality}, consisting of the following equations, 
\begin{eqnarray}
\frac{\partial S(x,t)}{\partial t} & = & D \frac{\partial^2 S(x,t)}{\partial x^2} + \mu N(x,t) - \mu  S(x,t) - \beta \frac{S(x,t) I(x,t)}{N(x,t)} 
\label{eq:sir1} \\
\frac{\partial I(x,t)}{\partial t} & = & D \frac{\partial^2 I(x,t)}{\partial x^2}+ \beta \frac{S(x,t) I(x,t)}{N(x,t)}  - \gamma I(x,t) -\mu I(x,t) \label{eq:sir2} \\
\frac{\partial R(x,t)}{\partial t} & = & D \frac{\partial^2 R(x,t)}{\partial x^2} + \gamma I(x,t) - \mu R(x,t) \label{eq:sir3}
\end{eqnarray}
where
\begin{equation}
S(x,t) + I(x,t) +R(x,t) = N(x,t) \label{eq:sir4}
\end{equation}
 $N(x,t)$ being the total population (density) at location $x$ and time $t$. 
 This 
 SIR model describes 
 a collection of individuals with a birth rate,  $\mu$, equal to the death rate, which is the same for all three populations, $S$, $I$ and $R$; 
 every individual in this system is born susceptible. All other terms refer to diffusion of the populations and the transition of individuals from one population to another. For instance the term $\gamma I(x,t)$ describes the rate at which infected individuals recover and are hence removed.
 
Adding (\ref{eq:sir1})--(\ref{eq:sir3}) we find
\begin{equation}
\frac{\partial N(x,t)}{\partial t} = D \frac{\partial^2 N(x,t)}{\partial x^2} \label{eq:sir5}
\end{equation}
This tells us that the overall population tends to an $x$ and $t$ independent distribution.  As a consequence we can assume a constant value for $N(x,t)=N$. This simplifies matters, and
we can replace $S(x,t)$ with $Ns(x,t)$ and similarly for $I(x,t)$ and $R(x,t)$. We then have the relationship
\begin{equation}
s(x,t) + i(x,t) + r(x,t) =1 \label{eq:sir6}
\end{equation}

\subsection{Uniform version of the SIR equations with vitality}
The three equations in this case are
\begin{eqnarray}
\frac{ds(t)}{dt} & = & \mu(1-s(t)) - \beta s(t)i(t) \label{eq:sir7} \\
\frac{di(t)}{dt} & = & \beta s(t) i(t) -(\gamma + \mu)i(t) \label{eq:sir8} \\
\frac{dr(t)}{dt} & = & \gamma i(t) - \mu r(t) \label{eq:sir9}
\end{eqnarray}
Equation (\ref{eq:sir3})  
is actually redundant, as we can also use Eq. (\ref{eq:sir6}) to determine $r(t)$. Furthermore, the variable $r(t)$ appears nowhere in Eqs. (\ref{eq:sir7}) and (\ref{eq:sir8}). 
If we assume as an initial condition $s(t) \approx 1$, with $r(0)=0$ and $i(0) =1-s(0)-r(0) $ very small, we see from (\ref{eq:sir8}) that the requirement for a growing infection is 
\begin{equation}
\beta > \gamma + \mu \label{eq:sir10}
\end{equation}

\subsection{Propagation of infection as a moving front: determination of the front's velocity}

\begin{figure}[htbp]
\begin{center}
\includegraphics[width=4in]{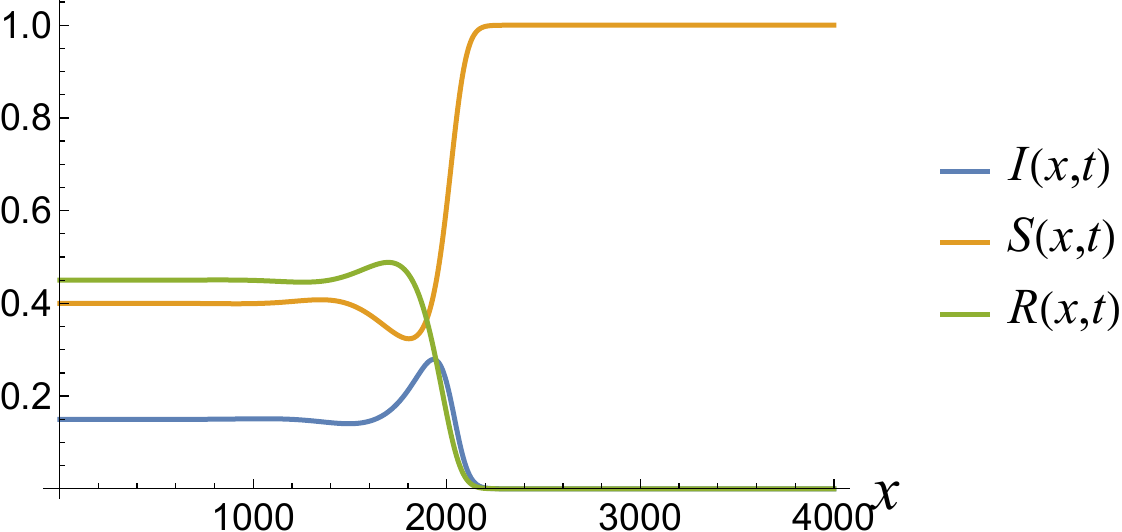}
\caption{A snapshot of the propagating infection front, in particular the values of $S(x,t)$, $I(x,t)$ and $R(x,t)$. In Eqs. (\ref{eq:sir1})--(\ref{eq:sir3}).  $N(x,t)$ has been set equal to 1, and the coefficients in the equation set are $D=300$, $\gamma=0.3$, $\mu=0.1$ and $\beta=1$. All functions translate at the same fixed speed from left to right. }
\label{fig:SIR_front1}
\end{center}
\end{figure}

As in the case of the SI equations of Section \ref{sec:noble}, when $\beta$ exceeds a threshold, in this case the inequality (\ref{eq:sir10}) being satisfied, 
and after transients have subsided, an initially localized region of infection will spread through the system as a soliton-like front. Figure \ref{fig:SIR_front1} is a snapshot of the front, which propagates by translation from left to right.

The infected population, $I(x,t)$, differs in its behavior from the the corresponding population in the Noble SI model, in that it does not relax to zero in the aftermath of the spreading front. Rather, as a result of the sustaining birth rate, the system relaxes to an ``endemic'' state in which a non-zero fraction of all individuals are
infected. The remainder is either uninfected and susceptible or previously infected and now recovered. The values of the coefficients  $\gamma$, $\mu$  and $\beta$ control the  distribution of the individuals among those three populations. Solving for a dynamically stable steady state solution to the constitutive uniform system equations, i.e., (\ref{eq:sir1})--(\ref{eq:sir3}) with left hand sides set equal to zero and $N(x,t)$ set equal to 1, we find
\begin{eqnarray}
S_{SS} & = & \frac{\gamma+\mu}{\beta} \label{eq:ss1} \\
I_{SS} & = & \frac{\mu( \beta - \gamma - \mu)}{\beta(\gamma+ \mu)} \label{eq:ss2} \\
R_{SS} & = & \frac{\gamma( \beta - \gamma - \mu)}{\beta(\gamma+ \mu)} \label{eq:ss3}
\end{eqnarray}
where the subscript $SS$ stands for ``steady state.''

To find the velocity of the moving front, we follow the steps in Section \ref{sec:velocity1}. We re-express $S(x,t)$ and $I(x,t)$ as in Eqs. (\ref{eq:outline20}) and (\ref{eq:outline21}). Then, defining \begin{eqnarray}
O_S &=& v\frac{d}{dx} +D \frac{d^2}{dx^2} + \mu \label{eq:newvel} \\
O_I& = & v\frac{d}{dx} +D \frac{d^2}{dx^2}  + \mu + \gamma \label{eq:newvel2}
\end{eqnarray}
the following equations will be satisfied for the proper choice of the front velocity, $v$,
\begin{equation}
O_S S_s(x) - \mu = -O_I I_s(x) = \beta S_s(x) I_s(x) \label{eq:newvel3}
\end{equation}
The expression for $R_s(x-vt)$ follows from the results above and 
\begin{equation}
S_s(x-vt) +I_s(x-vt) +R_s(x-vt) =N \label{eq:newvel4}
\end{equation}

\subsection{Behavior of the propagating infection front just above onset: uniform system analysis}

In the vicinity of onset, we write
\begin{equation}
\beta = \gamma + \mu + \epsilon \label{eq:sir11}
\end{equation}
with $\epsilon \ll 1$. Anticipating the behavior in the vicinity of onset, we make use of the re-parameterization in Section \ref{sec:SIthreshold}, writing 
\begin{eqnarray}
s(t) & = & 1+ \epsilon \sigma( \tau) \label{eq:sir12}  \\
i(t) & = & \epsilon \iota( \tau) \label{eq:sir13} \\
r(t) & = & \epsilon \rho( \tau) \label{eq:sir14}
\end{eqnarray}
where $\tau= \epsilon t$. 
In light of the sum rule (\ref{eq:sir6}), the following relationship now holds
\begin{equation}
\sigma( \tau) + \iota (\tau) + \rho(\tau) =0 \label{eq:sir15}
\end{equation}
(Note that, as defined, $\sigma < 0$.) 
We start by focusing on the equation for the evolution of the infected population, which can now be written
\begin{equation}
\epsilon^2 \frac{d \iota(\tau)}{d\tau} = \epsilon (\beta - (\gamma + \mu) ) \iota(\tau) + \epsilon^2 \beta \sigma(\tau)\iota(\tau)  \label{eq:sir16}
\end{equation}
Using (\ref{eq:sir11}) we transform (\ref{eq:sir16})  into
\begin{equation}
\frac{d \iota(\tau)}{d\tau} =  \iota(\tau) + \beta \iota(\tau) \sigma(\tau)  \label{eq:sir17}
\end{equation}
We now look at 
Eq. (\ref{eq:sir7}) 
under the substitutions (\ref{eq:sir12})--(\ref{eq:sir14}). Keeping leading-order-in-$\epsilon$ contributions to the right hand side of that equation we find
\begin{eqnarray}
\epsilon \frac{d \sigma(\tau)}{d\tau}&= & - \mu \sigma(\tau) - \beta \iota(\tau) \nonumber \\
& \rightarrow &  - \mu \sigma(\tau) - (\gamma + \mu) \iota(\tau) \label{eq:sir18}
\end{eqnarray}
Given the factor of $\epsilon$ in front of the derivative above, the leading-order-in-$\epsilon$ solution of this equation for   
$\sigma(\tau)$ is
\begin{equation}
\sigma(\tau) = - \frac{\gamma + \mu}{\mu} \iota(\tau) \label{eq:sir19}
\end{equation}
By a similar process we find, to leading order, 
\begin{equation}
\rho(\tau) = \frac{\gamma}{\mu} \iota(\tau)     \label{eq:sir20}
\end{equation}
These last two relations are entirely consistent with the sum rule (\ref{eq:sir15}). Making use of (\ref{eq:sir19}), Eq. (\ref{eq:sir17})
 becomes, to leading order in $\epsilon$, 
\begin{equation}
\frac{d \iota(\tau)}{d \tau} = \epsilon \iota(\tau) \left(1 - \frac{(\gamma+ \mu)^2}{\mu} \iota(\tau) \right) \label{eq:sir21}
\end{equation}
If we introduce the spatial dependence,  add diffusion and re-parameterize as in Section \ref{sec:SIthreshold}, we end up with the well-studied Fisher-KPP equation (see Eq. (\ref{eq:fkpp}) and accompanying discussion). 
This version of the equation is 
\begin{equation}
\frac{\partial \iota(\chi,\tau)}{\partial \tau} = D \frac{\partial \iota(\chi,\tau)}{\partial \chi^2} + \iota(\chi,\tau) \left(1-\frac{(\gamma+ \mu)^2}{\mu} \iota(\chi,\tau)\right) \label{eq:sir22}
\end{equation}
Once again, the velocity of propagation of the front can be established numerically by assuming an $\iota(\chi,\tau)$ of the form $\iota_s(\chi-v^{\prime}\tau)$, {so that the time derivative in (\ref{eq:sir22}) is replaced by $-v' \partial \iota/\partial \chi$,}
and then determining the value of $v^{\prime}$ for which the equation is satisfied.  The end result of this analysis is
\begin{equation}
v ^{\prime}= c_2 \sqrt{D}  \label{eq:sir23} 
\end{equation}
{where $c_2 \simeq 2$} 
in accord with well known results \cite{re:fisher37,re:kolmogorov37,Fife_Mcleod} {for the FKPP  equation}.
Returning to the original variables $x$ and $t$, we find for  the front's velocity in the vicinity of onset
\begin{eqnarray}
v &=& \epsilon^{1/2} v^{\prime} \nonumber \\
& = & 2 \sqrt{\epsilon D} \nonumber  \label{eq:SIRvfinal}
\end{eqnarray}
which exhibits the characteristic slowing down of the front velocity in the vicinity of onset.

\section{A model incorporating non-locality of the infection process}

Naether \textit{et. al.} \cite{re:naether08} 
proposed an epidemiological model that takes into account the mobility of the various populations and which results 
in a spatially extended 
infection process. Assuming that the the disease can be transmitted to a susceptible individual from members of the infected population in the neighborhood of that individual, and making use of a gradient expansion, the following {continuum} model is proposed,
\begin{eqnarray}
\frac{\partial S(x,t)}{\partial t} & = & - \beta I(x,t) S(x,t) - B S(x,t) \frac{\partial^2I(x,t)}{\partial x^2} \label{eq:N1} \\
\frac{\partial I(x,t)}{\partial t} & = & \beta I(x,t) S(x,t) + B S(x,t) \frac{\partial^2 I(x,t)}{\partial x^2} - \mu  I(x,t) \label{eq:N2} \\
\frac{\partial R(x,t)}{\partial t} & = & \mu I(x,t) \label{eq:N3}
\end{eqnarray}

\begin{figure}[htbp]
\begin{center}
\includegraphics[width=4in]{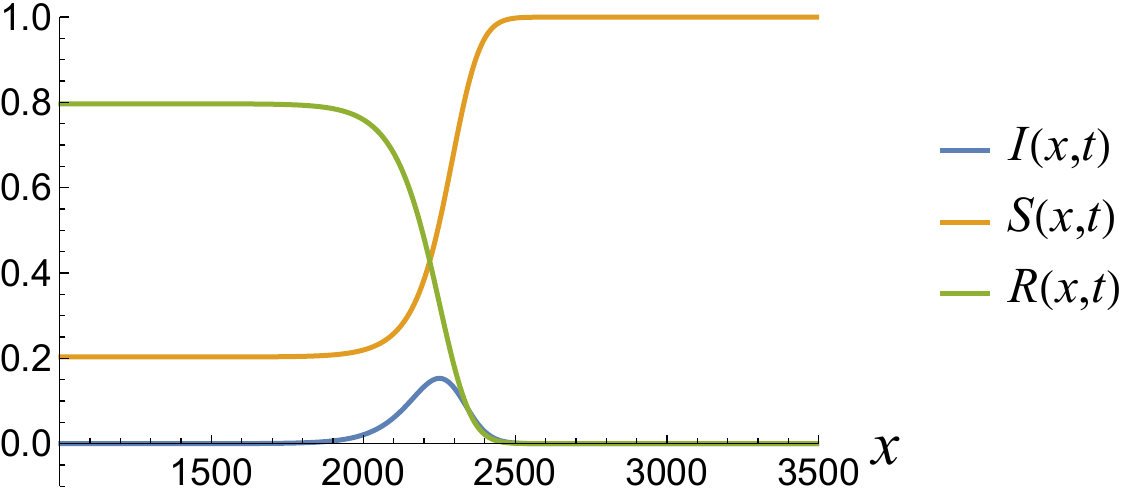}
\caption{A snapshot of the propagating front as generated by Eqs. (\ref{eq:N1})--(\ref{eq:N3}). The coefficients are as follows: $\beta=1$, $\mu =0.5$, $B=300$. The fronts move from left to right.  The higher rate of attrition of the infected population ($\mu=0.5$ here as opposed to $\mu=0.2$ in Fig. \ref{fig:fronts1}) leads to a larger surviving susceptible population.  }
\label{fig:Naetherfront}
\end{center}
\end{figure}

{For the uniform system, i.e., } in the absence of spatial derivatives on the right hand sides, the system above is just Noble's SI model for uniform populations (See Section \ref{sec:uniform}) with an added equation governing the removed individuals---the SIR equations without vitality. Numerical solution of Eqs. (\ref{eq:N1})--(\ref{eq:N3}) yields propagating fronts when, as in the case of the SI model, $\beta > \mu$.  Such  a propagating front is shown in Fig. \ref{fig:Naetherfront}

\subsection{Onset analysis}

The uniform equations {as noted,} are the just the SIR model without vitality. To investigate onset, we write $\mu = \beta(1-\epsilon)$  and make use of the parameterization in (\ref{eq:outline11}) and (\ref{eq:outline12}). Under the assumption {of a scaling solution}
the ``reduced" 
versions of the constitutive equations corresponding to (\ref{eq:outline17}) and (\ref{eq:outline18}) are
\begin{eqnarray}
\frac{\partial \sigma(\chi, \tau)}{\partial \tau} & = & - \beta \iota( \chi, \tau) \label{eq:Nonset1} \\
\frac{\partial \iota(\chi, \tau)}{\partial \tau} & = & \beta \sigma( \chi, \tau) \iota( \chi, \tau) + B \frac{\partial^2 \iota( \chi, \tau)}{\partial \chi^2}
\end{eqnarray}
We then assume soliton-like behavior, as embodied in 
\begin{eqnarray}
-v^{\prime} \sigma_s^{\prime}(\chi - v^{\prime} \tau) & = & - \beta \iota_s(\chi - v^{\prime} \tau) \label{eq:N4} \\
-v^{\prime}  \iota_s^{\prime}(\chi - v^{\prime}\tau) & = & \beta \sigma_s(\chi- v^{\prime} \tau) \iota_s (\chi- v^{\prime} \tau)+ B \iota_s^{\prime \prime}(\chi- v^{\prime} \tau) \label{eq:N5}
\end{eqnarray}
Finding, {numerically,} the value of $v^{\prime}$ that satisfies Eqs. (\ref{eq:N4}) and (\ref{eq:N5})  we discover that the velocity, $v^{\prime}$, in the vicinity of onset is given by
\begin{equation}
v =c \sqrt{\epsilon \beta B} \label{eq:N6}
\end{equation}
where $c \simeq 2.0$.

\section{The SEIR model}

An 
{extension} of the SIR model with vitality adds a population of \textit{exposed} individuals ($E$) {and an associated, additional time scale}. The members of this population {are assumed} to have taken up the pathogen through the infection process but 
are not symptomatic or able to pass it on, which is to say they are neither sick nor carriers. However, over the course of time, some fraction of the population does become truly infected. The rest are subject to the same other hazards as the susceptibles so have the same {(background)} death rate as the individuals in every other population. The equations governing this model are
\begin{eqnarray}
\frac{\partial S(x,t)}{\partial t} & = & D \frac{\partial^2 S(x,t)}{dx^2} + \mu N(x,t) - \mu S(x,t) - \frac{\beta S(x,t) I(x,t)}{N} \label{eq:SEIR1} \\
\frac{\partial E(x,t)}{\partial t} & = &D \frac{\partial^2 E(x,t)}{dx^2} + \frac{\beta S(x,t) I(x,t)}{N} - (\mu + a) E(x,t) \label{eq:SEIR2} \\
\frac{\partial I(x,t)}{\partial t} & = & D \frac{\partial^2 I(x,t)}{dx^2}  +aE(x,t) -(\mu + \gamma) I(x,t) \label{eq:SEIR3} \\ 
\frac{\partial R(x,t)}{\partial t} & = & D \frac{\partial^2 R(x,t)}{dx^2} + \gamma E(x,t) - \mu R(x,t) \label{eq:SEIR4}
\end{eqnarray}
with
\begin{equation}
N(x,t) = S(x,t) + E(x,t) + I(x,t) + R(x,t) \label{eq:SEIR5}
\end{equation}
The coefficient $a$ in Eqs. (\ref{eq:SEIR2}) and (\ref{eq:SEIR3}) quantifies the rate of transition of exposed individuals to the state of being infected.  It can readily be established that if $N(x,t)$ as given by (\ref{eq:SEIR5}) is spatially constant, then it remains unchanged under the action of Eqs. (\ref{eq:SEIR1})--(\ref{eq:SEIR4}). This means that $R(x,t)$  obeys the relationship 
\begin{equation}
R(x,t) =N-S(x,t) - E(x,t) -I(x,t) \label{eq:SEIR6}
\end{equation}
with the total population density, $N$, fixed in space and time. From here on 
% \sout{in}, 
we will set $N=1$. 

\subsection{Onset: the ODEs}

The analysis of onset begins with the consideration of the results of the fully mixed model, {i.e.,} the first order differential equations that ignore spatial dependence. Those equations are (see, e.g., (\ref{eq:SEIR1})--(\ref{eq:SEIR4}))
\begin{eqnarray}
\frac{dS(t)}{dt } & = & \mu(1-S(t)) - \beta S(t) I(t) \label{eq:SEIR7} \\
\frac{dE(t)}{dt} & = & \beta S(t) I(t) - (\mu + a) E(t) \label{eq:SEIR8} \\
\frac{dI(t)}{dt} & = & a E(t) -(\mu + \gamma)I(t) \label{eq:SEIR9} \\
\frac{dR(t)}{dt} & = & \gamma I(t) - \mu R(t) \label{eq:SEIR10}
\end{eqnarray}
We can ignore the last equation for two reasons. First, given the comments above, we can simply use the relation $S(t)+E(t) +I(t) +R(t) =1$ to solve for $R(t)$ given the values of the other populations. Second, the population $R(t)$ does not enter into the first three equations, which can be treated as a self-contained system. 

As a first step, we solve for fixed point solutions to (\ref{eq:SEIR8})--(\ref{eq:SEIR10}). Straightforward algebra leads to two {solutions;} the first fixed point corresponds to a population free of infection: $S=1, E=0, I=0, R=0$. The second set of solutions is
\begin{eqnarray}
S & = & \frac{(a+\mu ) (\gamma +\mu )}{a \beta } \label{eq:SEIR11} \\
E & = &\frac{\mu  (a \beta -(\mu +a) (\gamma +\mu ))}{a \beta  (a+\mu )} \label{eq:SEIR12} \\
I & = & \frac{\mu  (a \beta -(\mu +a) (\gamma +\mu ))}{\beta  (a+\mu ) (\gamma +\mu
   )} \label{eq:SEIR13} \\
   R & = & \frac{\gamma  (a \beta -(\mu +a) (\gamma +\mu ))}{\beta  (a+\mu ) (\gamma
   +\mu )} \label{eq:SEIR14}
\end{eqnarray}
The solutions above are consistent with realizable population values -- in that $E$, $I$ and $R$ are positive -- only if $\beta > (\mu+a)(\mu + \gamma)/a$. When this inequality is satisfied, the relations listed above describe the dynamically stable steady state of the equations (\ref{eq:SEIR11})--(\ref{eq:SEIR14}). 

Now, 
consider the implications 
when $\beta=(1+ \epsilon)  (\mu+a)(\mu + \gamma)/a$, with $\epsilon \ll 1$. To first order in $\epsilon$, the fixed point solutions are
\begin{eqnarray}
S & = & 1-\epsilon \label{eq:SEIR15} \\
E & = & \frac{\mu  \epsilon }{a+\mu } \label{eq:SEIR17} \\
I & = & \frac{a \mu  \epsilon }{(a+\mu ) (\gamma +\mu )} \label{eq:SEIR17} \\
R & = & \frac{a \gamma  \epsilon }{(a+\mu ) (\gamma +\mu )} \label{eq:SEIR18}
\end{eqnarray}

Next, we reparameterize the populations. Writing
\begin{eqnarray}
S(t) &=& 1+\epsilon s(t) \label{eq:SEIR19} \\
E(t) & = & \frac{\mu \epsilon}{\mu + a} e(t) \label{eq:SEIR20} \\
I (t)& = & \frac{a \mu \epsilon}{( \mu + a)( \mu + \gamma)} i (t)\label{eq:SEIR21}
\end{eqnarray}
with a similar---but unneeded---reparameterization for $R(t)$, we insert the right hand sides (\ref{eq:SEIR19})--(\ref{eq:SEIR21}) along with {$\beta = (1+ \epsilon)  (\mu+a)(\mu + \gamma)/a$} 
into Eqs. (\ref{eq:SEIR7})--(\ref{eq:SEIR9}). Expanding to second order in $\epsilon$, we find for the three relevant equations of motion,
\begin{eqnarray}
\epsilon \frac{ds(t)}{dt} & = & -\epsilon \mu( s(t) + i(t)) - \epsilon^2 \mu i(t) (1+s(t)) \label{eq:SEIR22} \\
\epsilon \frac{\mu}{\mu+ \alpha} \frac{de(t)}{dt} & = & \epsilon \mu (i(t) -e(t) ) + \epsilon^2 \mu i(t) (1+s(t)) \label{eq:SEIR23} \\
\epsilon \frac{a \mu}{(\mu +a)(\mu+ \gamma)} \frac{di(t)}{dt} & = & \epsilon \frac{a \mu}{(\mu + a)(\mu + \gamma)} (e(t) - i(t)) \label{eq:SEIR24}
\end{eqnarray}
Note that there is no term going as $\epsilon^2$ on the right hand side of (\ref{eq:SEIR24}). Under most circumstances we would be justified in ignoring all terms going as $\epsilon^2$. However, in this {case, with the addition of an additional subpopulation (E),} the higher-order 
 in $\epsilon$ contributions to the SEIR equations near onset prove to be quite {consequential, and the analysis of onset is more involved.} 

We start by looking at the the leading order terms in (\ref{eq:SEIR22})--(\ref{eq:SEIR24}). Dividing by $\epsilon$, the equations are 
\begin{eqnarray}
 \frac{ds(t)}{dt} & = & - \mu( s(t) + i(t))  \label{eq:SEIR25} \\
 \frac{\mu}{\mu+ \alpha} \frac{de(t)}{dt} & = & \mu (i(t) -e(t) ) \label{eq:SERI26} \\
 \frac{a \mu}{(\mu +a)(\mu+ \gamma)} \frac{di(t)}{dt} & = &\frac{a \mu}{(\mu +a)} (e(t) - i(t)) \label{eq:SEIR27}
\end{eqnarray}
Recasting the variables $s(t)$, $e(t)$ and $r(t)$ as the components of a column vector and simplifying notation, the relations above can be rewritten as
\begin{equation}
\frac{d}{dt} \left( \begin{array}{c} s(t) \\ e(t) \\ i(t) \end{array}  \right) =\left(
\begin{array}{ccc}
 -A & 0 & -A \\
 0 & -B & B \\
 0 & C & -C \\
\end{array}
\right) \left( \begin{array}{c} s(t) \\ e(t) \\ i(t) \end{array}  \right)  \label{eq:SEIR28}
\end{equation}
where 
\begin{eqnarray}
A& = & \mu \label{eq:SEIR29} \\
B& = & \mu + a \label{eq:SEIR30} \\
C& = & \mu + \gamma \label{eq:SEIR31}
\end{eqnarray}
To obtain the solution to this equation, we express the matrix on the right hand side as a ``spectral representation" in terms of its eigenvalues and associated projection operators.  The three eigenvalues are 0, $-A$ and $-B-C$. 
The projection operators satisfy ${ \bf P}_i \cdot { \bf P}_j = \delta_{i,j} { \bf P}_i$,
and are, in the order that the eigenvalues are listed,
\begin{eqnarray}
{ \bf P}_1 & = & \left(
\begin{array}{ccc}
 0 & -\frac{C}{B+C} & -\frac{B}{B+C} \\
 0 & \frac{C}{B+C} & \frac{B}{B+C} \\
 0 & \frac{C}{B+C} & \frac{B}{B+C} \\
\end{array}
\right) \label{eq:SEIR32} \\
{ \bf P}_2 & = & \left(
\begin{array}{ccc}
 1 & \frac{C}{-A+B+C} & -\frac{A-B}{-A+B+C} \\
 0 & 0 & 0 \\
 0 & 0 & 0 \\
\end{array}
\right) \label{eq:SEIR33} \\
{ \bf P}_3 & = & \left(
\begin{array}{ccc}
 0 & \frac{A C}{(B+C) (A-B-C)} & -\frac{A C}{(B+C) (A-B-C)} \\
 0 & \frac{B}{B+C} & -\frac{B}{B+C} \\
 0 & -\frac{C}{B+C} & \frac{C}{B+C} \\
\end{array}
\right) \label{eq:SEIR34}
\end{eqnarray}
The solution to (\ref{eq:SEIR28}) can then be written in the form
\begin{equation}
 \left( \begin{array}{c} s(t) \\ e(t) \\ i(t) \end{array}  \right) = \left({ \bf P}_1+ e^{-At} { \bf P_2} +e^{-(B+C)t} { \bf P}_3 \right) \cdot  \left( \begin{array}{c} s(0) \\ e(0) \\ i(0) \end{array}  \right) \label{eq:SEIR35}
\end{equation}
Note that at large values of $t$, the solution (\ref{eq:SEIR35}) becomes
\begin{equation}
{ \bf P}_1 \cdot  \left( \begin{array}{c} s(0) \\ e(0) \\ i(0) \end{array}  \right) = \left(\begin{array}{c}  -\frac{C e(0)}{B+C}-\frac{B i(0)}{B+C} \\\frac{C e(0)}{B+C}+\frac{B i(0)}{B+C}\\ \frac{C e(0)}{B+C}+\frac{B i(0)}{B+C} \end{array} \right) \label{eq:SEIR36}
\end{equation}
The solutions converge to $s(t)=-e(t)$ and $e(t) =i(t)$ but to limits determined by initial values and not to the dynamical fixed point, $s(t)=-1, e(t)=i(t)=1$. Approach to those fixed points is driven by the next-to-leading-order terms. 

Restoring the next-to-leading order terms, in generality we now have the set of equations
\begin{equation}
\frac{d}{dt}\left( \begin{array}{c} s(t) \\ e(t) \\ i(t) \end{array}  \right) =\left(
\begin{array}{ccc}
 -A & 0 & -A \\
 0 & -B & B \\
 0 & C & -C \\
\end{array}
\right) \left( \begin{array}{c} s(t) \\ e(t) \\ i(t) \end{array}  \right) + \epsilon \left( \begin{array}{c} -  \mu i(t) (1+s(t)) \\   \mu i(t) (1+s(t)) \\ 0 \end{array} \right) \label{eq:SEIR37}
\end{equation}
We then write
\begin{equation}
\left( \begin{array}{c} s(t) \\ e(t) \\ i(t) \end{array}  \right)  =  \left({ \bf P}_1+ e^{-At} { \bf P_2} +e^{-(B+C)t} { \bf P}_3 \right) \cdot \left( \begin{array}{c} s^{\prime}(t) \\ e^{\prime}(t) \\ i^{\prime}(t) \end{array}  \right) \label{eq:SEIR38}
\end{equation}
This allows us to cancel the leading order term on the right hand side of (\ref{eq:SEIR37}). The end result is 
\begin{equation}
\frac{d}{dt}\left( \begin{array}{c} s^{\prime}(t) \\ e^{\prime}(t) \\ i^{\prime}(t) \end{array}  \right)  = \epsilon \left({ \bf P}_1+ e^{At} { \bf P_2} +e^{(A+B)t} { \bf P}_3 \right) \cdot \left( \begin{array}{c} -  \mu i(t) (1+s(t)) \\   \mu i(t) (1+s(t)) \\ 0 \end{array} \right) \label{eq:SEIR39}
\end{equation}
Integrating up and making use of (\ref{eq:SEIR38}) we find
\begin{eqnarray}
\lefteqn{\left( \begin{array}{c} s(t) \\ e(t) \\ i(t) \end{array}  \right) } \nonumber \\ &=& \epsilon \int_0^t { \bf P}_1 \cdot \left( \begin{array}{c} -  \mu i(t^{\prime}) (1+s(t^{\prime})) \\   \mu i(t^{\prime}) (1+s(t^{\prime})) \\ 0 \end{array} \right) \, dt^{\prime}  \nonumber  \\ && + \epsilon \int_0^t e^{-A(t-t^{\prime})}{ \bf P}_2  \cdot \left( \begin{array}{c} -  \mu i(t^{\prime}) (1+s(t^{\prime})) \\   \mu i(t^{\prime}) (1+s(t^{\prime})) \\ 0 \end{array} \right) \, dt^{\prime} \nonumber \\ &&+ \epsilon \int_0^t e^{-(B+C)(t-t^{\prime})}{ \bf P}_3  \cdot \left( \begin{array}{c} -  \mu i(t^{\prime}) (1+s(t^{\prime})) \\   \mu i(t^{\prime}) (1+s(t^{\prime})) \\ 0 \end{array} \right) \, dt^{\prime} \label{eq:SEIR40}
\end{eqnarray}
Given that we expect the populations to evolve slowly, with a time constant of order $1/\epsilon$, we can find the leading order contributions to the second and third terms on the right hand side of (\ref{eq:SEIR40}). They yield 
\begin{equation}
(\frac{\epsilon}{A}{ \bf P}_2 +\frac{\epsilon}{B+C} { \bf P}_3) \cdot \left( \begin{array}{c} -  \mu i(t) (1+s(t)) \\   \mu i(t) (1+s(t)) \\ 0 \end{array} \right) \label{eq:SEIR41}
\end{equation}
Finally, we take the time derivative of the resulting equation. In light of (\ref{eq:SEIR41}) and the expectation of slow evolution of the densities, the time derivatives of the second and third terms in (\ref{eq:SEIR40}) will introduce a further factor of $\epsilon$ and can thus be discarded.   This leaves us with
\begin{eqnarray}
\frac{d}{dt}\left( \begin{array}{c} s(t) \\ e(t) \\ i(t) \end{array}  \right) &=& \epsilon{ \bf P}_1 \cdot \left( \begin{array}{c} -  \mu i(t) (1+s(t)) \\   \mu i(t) (1+s(t)) \\ 0 \end{array} \right) \label{eq:SEIR42}
\end{eqnarray}
Given Eq. (\ref{eq:SEIR32}), (\ref{eq:SEIR42}) becomes
\begin{equation}
\frac{d}{dt}\left( \begin{array}{c} s(t) \\ e(t) \\ i(t) \end{array}  \right)= \epsilon \frac{\mu(\mu+ \gamma)}{2 \mu + a + \gamma} \left( \begin{array}{c} - i(t) (1+s(t)) \\  i(t) (1+s(t)) \\  i(t) (1+s(t)) \end{array}\right) \label{eq:SEIR43}
\end{equation}
where we have used (\ref{eq:SEIR29})--(\ref{eq:SEIR31}). Given that the equations are consistent with $e(t)=i(t)=-s(t)$, we can recast the three resulting equations as follows
\begin{eqnarray}
\frac{ds(t)}{dt} & = & \epsilon \frac{\mu( \mu + \gamma)}{2 \mu + a + \gamma} s(t)(1+s(t)) \label{eq:SEIR44} \\
\frac{de(t)}{dt} & = & \epsilon \frac{\mu( \mu + \gamma)}{2 \mu + a + \gamma} e(t)(1-e(t)) \label{eq:SEIR45} \\
\frac{di(t)}{dt} & = & \epsilon \frac{\mu( \mu + \gamma)}{2 \mu + a + \gamma} i(t)(1-i(t)) \label{eq:SEIR46} 
\end{eqnarray}
These are all consistent with the Fisher-KPP equation as a description of the scaled densities in the vicinity of onset, as in the SIR equation with vitality -- recall that $s(t)$ is negative.  See Section  \ref{sec:SIR} for the full narrative  leading to that conclusion. 

\vspace{2 ex}

\section{Nonlinear infectivity constant and coexistence of two stable states}
All the models discussed so far assume that the likelihood of transmission goes as the product of the two population densities $S(x,t)$ and $I(x,t)$. This amounts to ignoring possible ``threshold'' effects, in which the rate of spread of the infection increases nonlinearly with the level of infection.  
Such effects can be encoded in an  $I$-dependence of the transmission coefficient $\beta$, so that $\beta \rightarrow \beta(I(x,t))$. As we shall see, a simple $I$ dependence leads in the modified SIR and SEIR models to the possibility of two dynamically stable steady states, one in which there is an endemic level of infection while the other is free of the disease. We start by considering a spatially uniform system. 
\subsection{The spatially uniform SIR system with an $I$-dependent $\beta$} \label{sec:nonlinear_uniform}

We return to 
Eqs. (\ref{eq:sir7})--(\ref{eq:sir9}) in which, now, $\beta$ takes the form
\begin{equation}
\beta(I) = \beta_0+ \alpha \frac{I}{N} \label{eq:newbeta}
\end{equation}
We then use the conservation of $s(t)+i(t)+r(t)$ to replace the sum by the constant value 1. Next, we choose a set of values for the coefficients in the resulting equations. Here, they are
\begin{eqnarray}
  \gamma = 0.3, \mu = 0.2, \beta_0 = 0.375, \alpha = 5
 \label{eq:rep_params}
\end{eqnarray} 
One important criterion for the choice of coefficients is that the value of $\beta_0$, the rate of infection at vanishingly small $I$, is below the threshold set by (\ref{eq:sir10}). in this case that threshold is 0.5. 

We then seek the stationary solutions of the first two of Eqs. (\ref{eq:sir7})--(\ref{eq:sir9}) (the solution of the fourth one following from $s(t)+e(t)+i(t)=1$) with $\beta$ replaced by $\beta(I)$ given by (\ref{eq:newbeta}). 
In this case one finds three sets of stationary solutions 
\begin{itemize}
\item {Solution 1:}
\begin{eqnarray}
s = 1 \label{eq:sol1s}
\end{eqnarray} 
\item {Solution 2:}
\begin{eqnarray}
s  =  \frac{\alpha  \mu +\sqrt{\left(\alpha  \mu -\beta _0 \beta _{\text{th}}\right){}^2-4
   \alpha  \mu  \beta _{\text{th}} \left(\beta _{\text{th}}-\beta _0\right)}+\beta _0
   \beta _{\text{th}}}{2 \alpha  \mu } \label{eq:sol2s} 
\end{eqnarray}
\item {Solution 3:}  
\begin{eqnarray}
s & = & \frac{\alpha  \mu -\sqrt{\left(\alpha  \mu -\beta _0 \beta _{\text{th}}\right){}^2-4
   \alpha  \mu  \beta _{\text{th}} \left(\beta _{\text{th}}-\beta _0\right)}+\beta _0
   \beta _{\text{th}}}{2 \alpha  \mu } \label{eq:sol3s} 
\end{eqnarray}
\end{itemize}
In the expressions above, $\beta_{\rm th}$ is the threshold value for the coefficient $\beta$, in the case  of an $I$-independent $\beta$ given by
\begin{equation}
\beta_{\rm th} = \gamma+ \mu \label{eq:betath}
\end{equation}
(See Eq. (\ref{eq:sir10}) and accompanying text.) In all three cases above, the stationary solution for $i$ and $r$ are given by
\begin{eqnarray}\
i &= & \mu \frac{1-s}{\beta_{\rm th}}  \label{eq:newieq} \\
r & = & \gamma \frac{1-s}{\beta_{\rm th}} \label{eq:newreq}
\end{eqnarray}
Finally, we recover the steady state solution for $r$ from the sum rule above.

One absolute requirement that must be satisfied by the steady state solutions for the sub-populations is that they are real and positive. In the case of the second and third
solutions listed above, reality of the population values 
requires a positive argument 
in the square roots in (\ref{eq:sol2s}) and (\ref{eq:sol3s}). Coupled with the demand for positive sub-populations we are led to  leads to the following inequality,
\begin{equation}
\alpha > \frac{2 \sqrt{\mu ^2 \beta _{\text{th}}^3 \left(\beta _{\text{th}}-\beta _0\right)}-\mu 
   \left(\beta _0-2 \beta _{\text{th}}\right) \beta _{\text{th}}}{\mu ^2} \label{eq:alphainequality}
\end{equation}

Next,  we determine the stability of those solutions. At this point, we replace coefficients by the {representative} values listed 
above.
Expanding about the three solutions, we find for the eigenvalues of the linear stability matrix of the  solutions,
\begin{itemize}
\item {Solution 1:} $ \lambda_1=-0.2, \lambda_2 = -0.125 $
\item{Solution 2:}  $\lambda_1=-0.182396, \lambda_2=0.120832 $
\item{Solution 3:} $\lambda_1=-0.166874 + 0.397813 i, \lambda_2=-0.166874 - 0.397813  i $
\end{itemize}
We see that the eigenvalues of the linear stability matrix for Solution 1, which corresponds to an uninfected population, are all negative; the uninfected state is  stable with respect to the introduction of an infection, assuming the initial rate of infection is sufficiently low. Additionally, Solution 3, in which there is an endemic 
infection rate, is also stable with respect to small perturbations, in that the real part of the eigenvalue is negative. On the other hand in the case of Solution 2---in which the rate of infection lies between 0 and the steady state value in Solution 3---one of the eigenvalues of the linear stability matrix is positive, so it is dynamically unstable.

\begin{figure}[htbp]
\begin{center}
\includegraphics[width=3in]{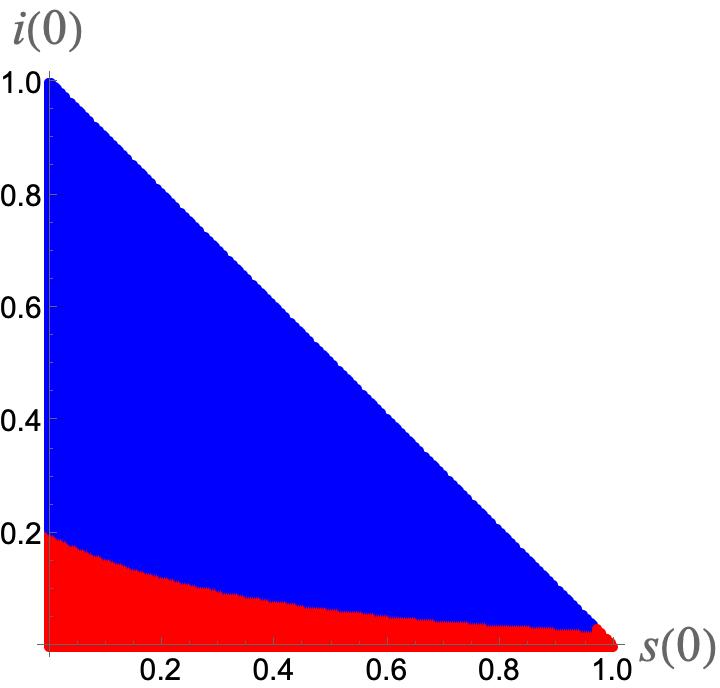}
\caption{The basins of attraction for the SIR model with modified $\beta$ as given by (\ref{eq:newbeta}), coefficients $\gamma$, $\mu$ and $\beta_0$ fixed according to (\ref{eq:rep_params}) and $\alpha=5$. The axes are the initial values of $s(t)$ and $i(t)$, which  both range from 0 to 1 subject to the restriction that $s(0)+i(0) \leq 1$. The value of $r(0)$ is given by $r(0)=1-s(0)-i(0)$. In the blue region the fixed point to which the system converges is endemically infected, corresponding to solution 3 for $s$, with $i$ and $r$ given by (\ref{eq:newieq}) and (\ref{eq:newreq}).  In the red region the limiting values correspond to the uninfected state: $s=1$, $i=r=0$.  }
\label{fig:basins1}
\end{center}
\end{figure}

\begin{figure}[htbp]
\begin{center}
\includegraphics[width=3in]{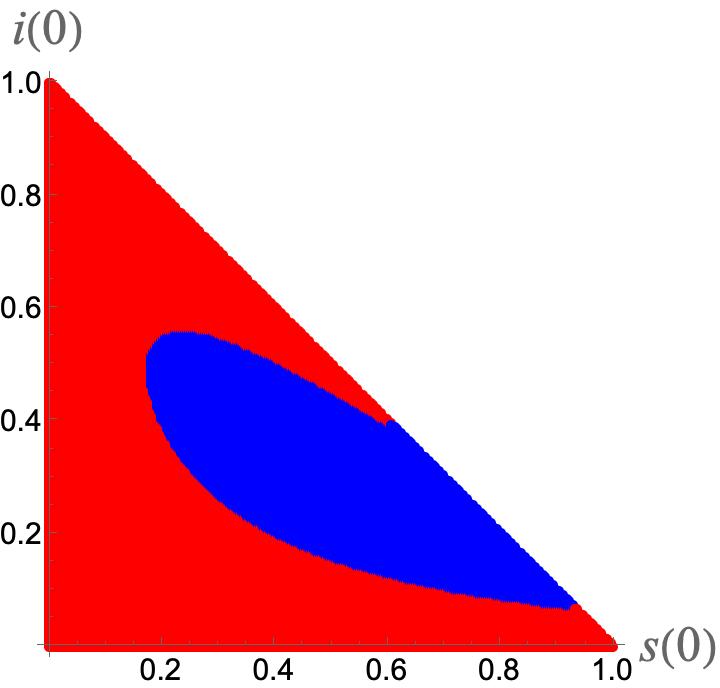}
\caption{The basins of attraction with all coefficients the same as in Fig. \ref{fig:basins1}, with the exception of $\alpha$, which is now 3. Again, the red region is corresponds to an uninfected limiting state and the blue region to the limit of an endemic infection. }
\label{fig:basins2}
\end{center}
\end{figure}

We can explore the basins of attraction of the two stable stationary states by solving the full set of spatially uniform equations, (\ref{eq:SEIR7})--(\ref{eq:SEIR10}). Figures \ref{fig:basins1} and \ref{fig:basins2} show what we find for coefficients $\gamma$, $\mu$ and $\beta_0$ fixed according to 
(\ref{eq:rep_params}),
and two different values of $\alpha$. Figure \ref{fig:basins1} displays the basins of attraction when $\alpha=5$. 
Figure \ref{fig:basins2} displays the basins of attraction when $\alpha=3$, close to the minimum value of 2.815 allowed by the requirement of reality and positivity of the dynamically stable fixed points, as expressed in (\ref{eq:alphainequality}). As $\alpha$ approaches that value the blue basin of attraction continuously shrinks to zero. 

\pagebreak

\subsection{Two regions with different levels of infection in the SIR model with an $I$-dependent $\beta$} 

The results in Section \ref{sec:nonlinear_uniform}  point to the prospect of competition---and possibly  coexistence---between  two regions with different levels of infection, specifically an uninfected region and a region sustaining an endemic infection.  

To see how this can come about, we turn our focus to the full set of spatio-temporal eqquations, (\ref{eq:sir1})--(\ref{eq:sir3})  with the $I(x,t)$-dependent $\beta$ given by (\ref{eq:newbeta}), $I$ on the right hand side being replaced by $I(x,t)$. We can replace $N(x,t)$ by a constant value, which we set equal to 1, and  we have the sum rule
\begin{equation}
S(x,t) +I(x,t) + R(x,t) =1 \label{eq:seirsumrule}
\end{equation}
which means we need only concern ourselves with the first two equations. 

\begin{figure}[htbp]
\begin{center}
\includegraphics[width=4in]{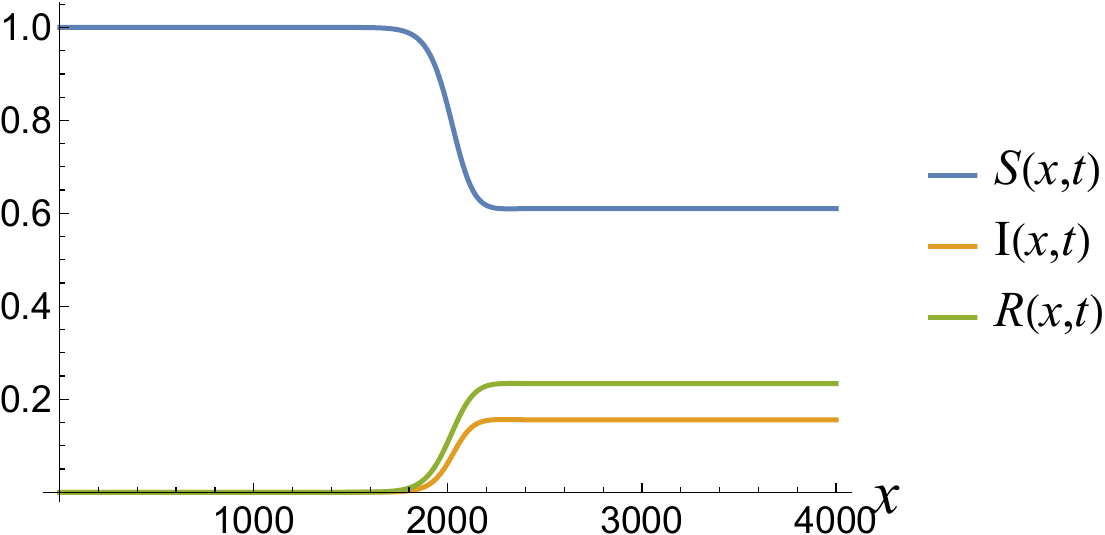}
\caption{Snapshot of the sub-populations in two regions. On the left the population is infection free and on the right, there is infection at an endemic level. The coefficient $\alpha$ in (\ref{eq:newbeta}) has been set equal to 2.9, and the front between the two regions propagates to the right, with a velocity 0.66776, so the system is evolving towards a uniform uninfected state.  }
\label{fig:coexistence}
\end{center}
\end{figure}

\begin{figure}[htbp]
\begin{center}
\includegraphics[width=4in]{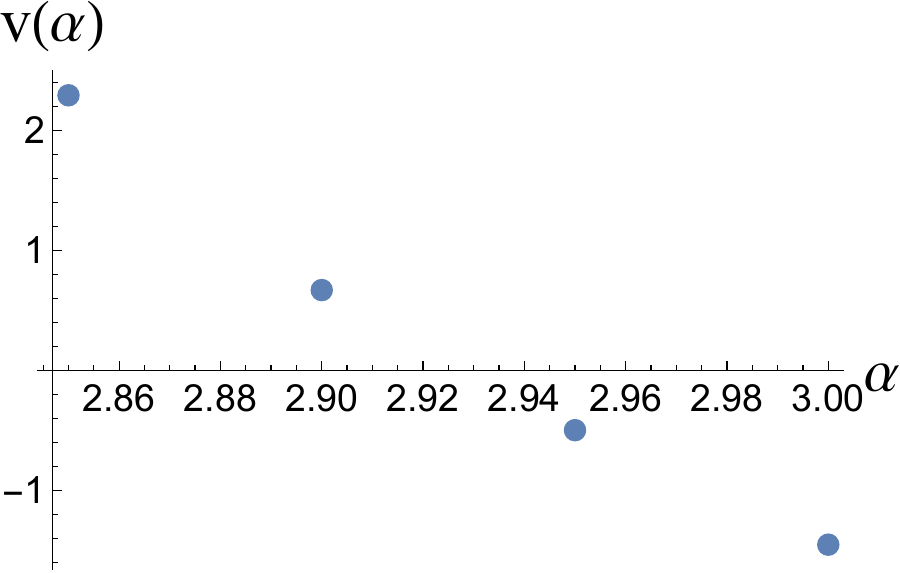}
\caption{The velocity $v(\alpha)$ of the propagation of the fronts, as shown in Fig. \ref{fig:coexistence}, as a function of the coefficient $\alpha$ in (\ref{eq:newbeta}). The value of $\alpha$ for which the fronts do not propagate either to the left, as they do for higher values of $\alpha$, or to the right, as they do for lower values,  is $ \alpha_0 \approx 2.927$.  }
\label{fig:vplot}
\end{center}
\end{figure}

Numerical simulations using these equations with parameters $\gamma$, $\mu$ and $\beta_0$ set according to (\ref{eq:rep_params})
and $\alpha$ in the immediate vicinity of 3 yield two regions separated by a compact front, as shown in Fig. \ref{fig:coexistence}. 
The front can propagate in either direction, depending on the choice for $\alpha$. For $\alpha< \alpha_0 \approx 2.9267$ the front propagates to the right, corresponding to an evolution to a uniform uninfected state, while if $\alpha$ is greater than that transition value the propagation is to the left, and the system ends up converting to a uniformly endemic state. The front remains compact as $\alpha$ is assigned values in the immediate vicinity of $\alpha_0$. When $\alpha=\alpha_0$, the front is stationary, and we have perfect coexistence of the regions. 
When the fronts propagate, they do so in a solitonic manner. This can be verified, and the dependence of the velocity on the value of $\alpha$ determined, by testing the assumption that $S(x)=S_s(x-vt)$ and similarly for $E(x)$, $I(x)$ and $R(x)$. Then the SIR equations reduce to
\begin{eqnarray}
-v\frac{d S_s(x)}{d x} & = & D \frac{d^2 S_s(x)}{dx^2} + \mu - \mu S_s(x) - \beta(I_s(x)) S_s(x) I_s(x) \label{eq:SIR1n} \\
-v\frac{d I_s(x)}{d x} & = & D \frac{d^2 I_s(x)}{dx^2}  +aE_s(x) -(\mu + \gamma) I_s(x) \label{eq:SIR2n} \\ 
-v\frac{d R_s(x)}{d x} & = & D \frac{d^2 R_s(x)}{dx^2} + \gamma E_s(x) - \mu R_s(x) \label{eq:SIR3n}
\end{eqnarray}
where we have set the total population, $N_s(x) =1$. We then solve the original set of spatio-temporal equations (\ref{eq:sir1})--(\ref{eq:sir3}) numerically, take a snapshot of the system, such as the one shown in Fig. \ref{fig:coexistence}, and check whether there is a value of $v$ for which Eqs. (\ref{eq:SIR1n})--(\ref{eq:SIR3n}) are all satisfied. To satisfactory numerical tolerance (agreement to five significant figures in the velocities) we find this to be the case for coefficients as given by (\ref{eq:rep_params})
with $\alpha$ varying from  2.85 to 3. The dependence of $v$ on $\alpha$ is shown in Fig. \ref{fig:vplot}. 

\subsection{The case of the SEIR model with an $I$-dependent $\beta$}

The analysis of an SEIR model in which the coefficient $\beta$ has the $I$ dependence (\ref{eq:newbeta}) is similar, and in some cases identical, to the discussion above. The homogeneous system admits three fixed points. The values of $s$ at those fixed points are given by  (\ref{eq:sol1s})--(\ref{eq:sol3s}), the one change being $\beta_{\rm th}$, used which in this case is 
\begin{equation}
\beta_{\rm th} = \frac{(\mu+a)(\mu + \gamma)}{a}  \label{eq:nbetath}
\end{equation}
Given the fixed point value of $s$, the populations $e$, $i$ and $r$ are
\begin{eqnarray}
e & = & \frac{\mu(1-s)}{\mu+a}  \label{eq:fixede} \\
i & = & \frac{ \mu (1-s)}{\beta_{\rm th}}  \label{eq:fixedi} \\
r & = & \frac{\gamma(1-s)}{\beta_{\rm th}}
\end{eqnarray}
The stability properties of the three fixed points are the same as in the case of the SIR model. The basins of attraction can be determined, with the complication that they are now three dimensional manifolds. In order for the system to admit of an uninfected stable fixed point, it is necessary to adjust $\beta_0$ to be less than $\beta_{\rm th}$. With $\gamma=0.3$, $\mu=0.2$ and $a=0.4$, $\beta_{\rm th}=0.75$. Choosing $\beta_0=0.675$ we find that for $\alpha \approx 5.0533$ a state with endemic infection coexists with an uninfected state, while for $\alpha$ slightly different than that value the front between those states propagates and one of the states grows at the expense of the other.

There are a number of interesting aspects to the possibility of a stationary kink, since there may be real world examples in which two neighboring regions may "coexist" with disparate infection levels. In fact, the existence of a stationary kink, which requires precise tuning of parameters ($\alpha$, for fixed $\beta_0, \gamma, \mu, a$), and an advancing front with even slight de-tuning, is analogous to a first-order transition in thermodynamics. For example, when the pressure $P$ is {\bf on} the vapor pressure curve $P_\sigma(T)$  with $T < T_c,$ the critical temperature, a liquid can coexist with its vapor with any volume fractions for the two phases. In principle, with $P - P_\sigma(T) = 0\pm$, all of the material will convert to one phase or the other in equilibrium occupying the system. If the pressure is adjusted rapidly, the interface separating the liquid and its vapor (a kink solution) will move one way or the other  \cite{Cross} until one phase fully occupies the system. This dynamical behavior can be seen in SEIR compartmental models  as discussed here, with $\alpha_0 - \alpha_c$ playing the role of the appropriate chemical potential difference (or magnetic field) the context of condensed matter. Appendix \ref{sec:front}  outlines the analysis of such a model.

\section{Summary}

We have examined several extensions to compartment models that incorporate spatial dependence of the relevant variables so as to model either individual mobility or disease spreading through contact between individuals at different geographical locations. All models considered here can be written as simple reaction-diffusion systems, and we find that stable propagating front solutions generically exist in certain ranges of model parameters that depend on the specifics of each model. In that case, we also find the velocity of the propagating front. These results potentially provide an additional avenue for obtaining model parameters from empirical infection data, independently from estimates of the various rates from single compartment data. 

The various models studied have to be considered mean-field approximations in which local statistical fluctuations in the subpopulations have been neglected. This is only realistic in cases in which all the subpopulations are large so that the various rates of interaction introduced in the models can be meaningfully defined. While this may well be the case for some diseases, it is expected to fail for the SARS-CoV-2 pandemic. The rates of infection are quite low relative to the size of the population,
and transmission and spread is expected to be dominated by large fluctuations. As a consequence, one would not expect, and indeed does not observe in the empirical data, compact fronts separating regions at different stages of infection \cite{re:nyt22,re:li22}. Current extensions of compartment models therefore involve networks with a very large number of nodes to analyze the observed effects of subpopulation heterogeneity. Whether such heterogeneity can be modeled via stochastic extensions of the compartment models is currently under study.

%\bibliography{references_dj}   


\appendix

\section{The FKPP model}
\label{sec:fkpp}

The one dimensional Fisher-Kolmogorov–Petrovsky–Piskunov 
equation \cite{re:fisher37,re:kolmogorov37}  determines the evolution of a scalar field $u(x,t)$ which obeys the equation,
\begin{equation}
\frac{\partial u}{\partial t} = D \frac{\partial^{2} u}{\partial x^{2}} +
ru(1-u),
\label{eq:fkpp}
\end{equation}
where both $D > 0$ and $r >0$ are model parameters. The equation was originally introduced to describe the spatial propagation of an advantageous gene, but given the generality of the reaction term (the last term in the right hand side of Eq. (\ref{eq:fkpp}){\color{red})}, this equation has been used to describe front propagation in many different contexts in Physics, Chemistry, and Biology, including Ecology, Physiology, combustion, and solidification. The equation  as written has two fixed points $u = 0$, and $u = 1$. The fixed point $u = 0$ is linearly unstable, whereas the fixed point $u = 1$ is linearly stable to perturbations of the  form $u(x,t) = 1 + \epsilon \cos(qx)$, $\epsilon \ll 1$ while $\sigma = -r  + D q^{2} < 0$. The equation has front solutions $u(x,t) = f(x-vt)$ that satisfy the boundary conditions $u(x \rightarrow - \infty) = 0$, $u(x
\rightarrow \infty) = 1$ so that the front advances at constant speed $v = 2 \sqrt{Dr}$.

\section{Motion of a one-dimensional interface separating two different phases} \label{sec:front}

We start with an equation that closely resembles the Fisher-Kolmogorov–Petrovsky–Piskunov (FKPP) \cite{re:fisher37,re:kolmogorov37} equation in (\ref{eq:fkpp}). 
\begin{equation}
\frac{\partial u}{\partial t} = D \frac{\partial^2 u}{\partial x^2} +ru(1-u^2) \label{eq:front1}
\end{equation}
This equation, unlike the FKPP equation, supports a stationary front. To see that this is so, set the left hand side of (\ref{eq:front1}) equal to zero, and insert the solution
\begin{equation}
u_s(x) = \tanh( B(x-x_0))  \label{eq:front2}
\end{equation}
into the new equation. Taking the second derivative of $u_s(x)$, we find
\begin{eqnarray}
D\frac{d^2u_s(x)}{dx^2} & = & -2DB^2 \tanh(B(x-x_0)) {\mathop{\rm sech}}^2(B(x-x_0)) \nonumber \\ &  = & -2DB^2 u_s(x)(1-u_s(x)^2) \label{eq:front3}
\end{eqnarray}
The stationary solution of (\ref{eq:front1})---i.e. $u$ is independent of $t$---is thus satisfied if
\begin{equation}
B=\sqrt{\frac{r}{2D}} \label{eq:front4}
\end{equation}

\begin{figure}[htbp]
\begin{center}
\includegraphics[width=3in]{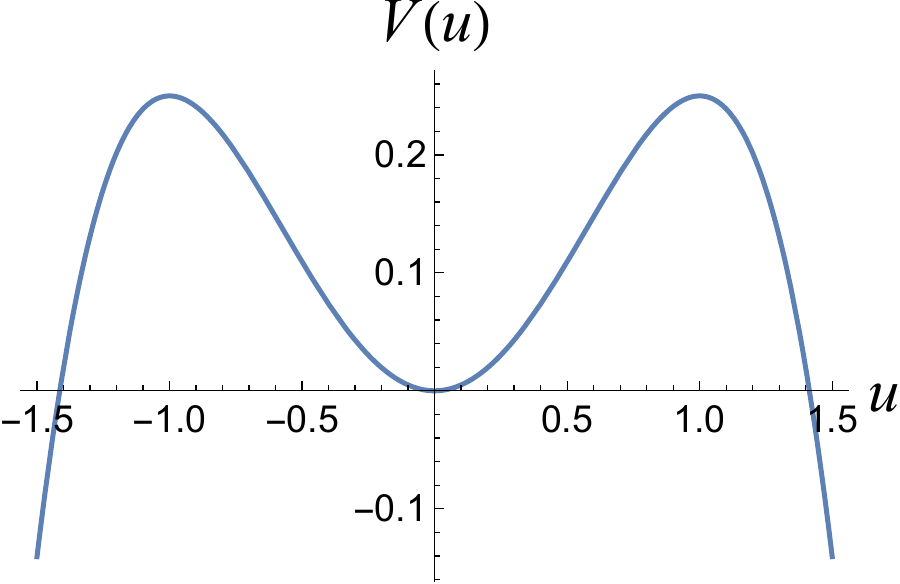}
\caption{The potential function $V(u)$ as given by (\ref{eq:front6}), with $r=1$. }
\label{fig:potfun}
\end{center}
\end{figure}

There are two instructive physical interpretations of the stationary version of Eq. (\ref{eq:front1}). First, thinking of $u$ as a position variable and $x$ as a time variable, the equation
\begin{equation}
D \frac{d^2u(x)}{dx^2} =-ru(x)(1-u(x)^2) \label{eq:front5}
\end{equation}
is the equation of motion of a non-relativistic point particle with a potential energy of the form 
\begin{equation}
V(u)=r \left(\frac{u^2}{2} -\frac{u^4}{4} \right) \label{eq:front6}
\end{equation}
Second, thinking $u(x)$ as a scalar property, like the local magnetization along an easy axis, of a system with spatial variation in one dimension, Eq. (\ref{eq:front5}) is the minimization equation for the free energy of a system, if that free energy takes the form
\begin{eqnarray}
\lefteqn{\int_{-\infty}^{\infty}\left[\frac{1}{2}\left( \frac{du(x)}{dx}\right)^2  - r \left( \frac{u(x)^2}{2}-\frac{u(x)^4}{4} \right) \right] dx} \nonumber \\
& = & \int_{-\infty}^{\infty}\left[\frac{1}{2} \left(\frac{du(x)}{dx} \right)^2 -V(u(x)) \right] dx \label{eq:front7}
\end{eqnarray}
This is an example of a Ginzburg-Landau-like free energy \cite{Landau,Ginzburg1}. A graph of $V(u)$ is shown in Fig. \ref{fig:potfun}.

From the dynamical viewpoint, the solution (\ref{eq:front2}), with (\ref{eq:front4}), describes the motion of a point mass  that starts out at rest on the leftmost peak at $u=-1$ and moves through the valley, eventually coming to rest on the rightmost peak at $u=1$. From the statistical mechanical viewpoint, the solution describes the profile of a domain wall separating a region in which the system is uniform with a  negative magnetization corresponding to a minimum of $-V(u)$ and a region of positive magnetization corresponding to the other minimum of that function. 

Before proceeding to the case of a propagating front solution to Eq. (\ref{eq:front1}), let's note the arbitrariness of the term $x_0$ in the solution (\ref{eq:front2}), which reflects the translational invariance of the solution. If we were to change $x_0$ to $x_0 + \delta x$. The solution, to first order in $\delta x$, is 
\begin{equation}
u_s(x) -u^{\prime}_s(x) \delta x \label{eq:front8}
\end{equation}
This new function is still a solution to the stationary equation, which means substituting (\ref{eq:front8}) into (\ref{eq:front5}) we find no term going as $\delta x$. The function $u_s^{\prime}(x)$ is an example of what is known as a translation mode \cite{Coleman,translation}.

Now, suppose we change the term $V(u)$ from its original form to $V(u)+ \Delta V(u)$. Anticipating that this change induces motion in the stationary front and that to lowest order in $\Delta V(u)$ the moving front has the same general form as the stationary one, we replace $u(x)$ with $u_s(x-vt)$ with $v$ small. Substituting this in Eq. (\ref{eq:front1}), expanding to first order in $\Delta V(u)$ and $u$ we obtain
\begin{equation}
-v u_s^{\prime}(x-vt) = -\Delta V^{\prime}(u_s(x-vt)) - \delta V^{\prime} \label{eq:front9}
\end{equation}
The term $\delta V^{\prime}$ consists of first order changes in $V^{\prime}(u)$  in $u(x-vt) -u_s(x-vt)$. This term can be shown to be mathematically orthogonal to $u^{\prime}_s(x-vt)$.  Multiplying both sides of this equation by $u_s^{\prime}(x-vt)$ and integrating we find
\begin{eqnarray}
v&=& \frac{\int_{-\infty}^{\infty} \Delta V^{\prime}(u_s(x-vt))u_s^{\prime}(x-vt) dx}{\int_{-\infty}^{\infty} \phi_s^{\prime \, 2}(x-vt)dx} \nonumber \\ & = & \frac{\Delta V(x=\infty) - \Delta V(x=-\infty)}{\int_{-\infty}^{\infty} \phi_s^{\prime \, 2}(x-vt)dx}
\end{eqnarray}
The velocity of propagation of the front depends on the difference between the two limiting values of the change in the local free energy term. 

\end{document}